\documentclass[twocolumn,preprintnumbers,amsmath,amssymb]{revtex4-1}
\usepackage{epsfig}
\usepackage{color}
\usepackage{amsmath}
\usepackage{float}
\usepackage{amsfonts}
\usepackage{amssymb}
\usepackage{multirow}
\usepackage{graphicx,stackengine}
\usepackage{subfig}
\usepackage[percent]{overpic}
\usepackage[english]{babel}
\usepackage{float}
\usepackage{array}
\usepackage{pst-plot}
\usepackage[version=4]{mhchem}
\usepackage{lipsum}
\usepackage{verbatim}
\usepackage[hidelinks]{hyperref} % add hypertext capabilities
\usepackage{tikz}
\usepackage{natbib}
\usetikzlibrary{calc,fadings}
\tikzfading[name=fade l,left color=transparent!100,right color=transparent!0]
\tikzfading[name=fade r,right color=transparent!100,left color=transparent!0]
\tikzfading[name=fade d,bottom color=transparent!100,top color=transparent!0]
\tikzfading[name=fade u,top color=transparent!100,bottom color=transparent!0]

\begin{document}
\title{Understanding How Synthetic Impurities Affect Glyphosate Solubility and Crystal Growth Using Free Energy Calculations and Molecular Dynamics Simulations}

\author{Alejandro Castro$^{1,+}$, Ignacio Sanchez-Burgos$^{2,+}$, Nuria H. Espejo$^{1,3}$, Adiran Garaizar$^{3,*}$, Giovanni Maria Maggioni$^{4,*}$, Jorge R. Espinosa$^{1,2,5*}$}
\affiliation{
[1] Department of Physical Chemistry, Universidad Complutense de Madrid, Av. Complutense s/n, Madrid 28040, Spain \\
[2] Yusuf Hamied Department of Chemistry, University of Cambridge, Lensfield Road, Cambridge CB2 1EW, UK \\ 
[3] Data Science, Bayer AG, Alfred-Nobel-Straße 50,
40789 Monheim am Rhein, Germany \\
[4] Crop Protection Innovation - Bayer AG, Kaiser-Wilhelm-Allee 1, 51373 Leverkusen, Germany \\
[5] Multidisciplinary Institute, Complutense University of Madrid, Paseo Juan XXIII, 1, Madrid 28040, Spain \\
* = To whom correspondence should be sent.
email: adiran.garaizarsuarez@bayer.com , john.maggioni@bayer.com , jorgerene@ucm.es\\
$^+=$These authors contributed equally}

\date{\today}

\begin{abstract}
    
\begin{center}
    \textcolor{black}{\textbf{ABSTRACT}}
\end{center}

Glyphosate, the most widely used herbicide worldwide, crystallizes through complex intermolecular interactions that are strongly influenced by synthesis-derived impurities. Understanding this process at the molecular scale is critical for optimizing production, ensuring product quality, and assessing environmental impact. Here, we employ direct coexistence molecular dynamics simulations and free energy calculations to elucidate how glycine---a prevalent synthesis byproduct---modulates glyphosate solubility and crystal growth in aqueous solutions. Our simulations identify two major mechanisms by which glycine hinders crystallization. First, direct coexistence simulations show that glycine preferentially adsorbs at crystal surfaces, hindering glyphosate attachment and slowing growth. Second, free energy calculations demonstrate that glycine enhances glyphosate solubility, reducing the supersaturation driving force to incorporate into the crystal phase. Experimental measurements corroborate our predictions, confirming both enhanced solubility and reduced crystallization kinetics in glycine-bearing systems. These findings establish that glycine---typically considered an inert impurity---actively disrupts glyphosate crystallization by promoting its dissolution. More broadly, this integrated computational–experimental approach highlights the power of molecular simulations to disentangle impurity effects, interfacial phenomena, and solution thermodynamics in crystallization, providing molecular-level insights for optimizing industrial protocols and predicting agrochemical behavior under relevant environmental conditions.
\end{abstract}

\keywords{Glyphosate, crystallization, molecular dynamics simulations, solubility, glycine \\ $^+=$These authors contributed equally
}

\maketitle

%\pagebreak

\section*{Introduction} \label{sec:intro}

Glyphosate (N-phosphonomethylglycine), the world's most widely used herbicide \cite{duke2008glyphosate}, inhibits 5-enolpyruvylshikimate-3-phosphate synthase \cite{schonbrunn2001interaction}, an enzyme essential for biosynthesis of aromatic amino acids including phenylalanine, tyrosine, and tryptophan in a wide range of plants \cite{herrmann1999shikimate}. The absence of this shikimate pathway in animals confers glyphosate selective toxicity toward plants \cite{10.3389/fenvs.2016.00028}.
%Due to its effectiveness, glyphosate is the most widely used herbicide globally, as its nonselective activity inhibits ESPS synthase activity in a wide range of plants, which is crucial to vegetal life\cite{10.3389/fenvs.2016.00028}. It is commercialized under the brand Roundup$^{\circledR}$ by BAYER and has played a central role in modern agriculture, being termed a "once-in-a-Century" herbicide\cite{duke2008glyphosate}.
Structurally, glyphosate is a glycine analog bearing a phosphonomethyl substituent at the nitrogen atom. Its molecular architecture incorporates carboxylic acid, amine, and phosphonic acid functionalities, conferring it an amphoteric character with multiple ionization states. Under environmental conditions (pH 6–8), glyphosate exists predominantly as a dianionic or trianionic specie \cite{Abate01011996}. Its high polarity and multiple ionizable groups drive strong interactions with both organic matter \cite{sprankle1975adsorption} and soil minerals \cite{sprankle1975rapid}, influencing its environmental fate and strong bioavailability.

Crystallization represents a critical purification step in glyphosate synthesis which determines product quality \cite{zhou2012study}. While impurities do not directly affect its herbicidal activity, they are essential considerations for regulatory compliance and quality control \cite{ATSDR2020Glyphosate}. Impurities significantly influence crystallization kinetics through mechanisms that remain poorly characterized \cite{tosylate}. This complexity arises from glyphosate's pH-dependent zwitterionic structure \cite{Sandercock01011997}, extensive hydrogen-bonding network \cite{knuuttila1979crystal}, and polymorphic behavior under varying pH and pressure conditions \cite{wilson2023discerning,messer2005ph}. Molecular-level understanding of its solubility, nucleation rate and crystal growth \cite{tejedor} is therefore essential for optimizing crystallization protocols and elucidating impurity effects on crystal formation \cite{WINDOM2022120154}. In that sense, molecular dynamics (MD) simulations represent a powerful tool for obtaining relevant atomistic insights into these processes.

Here, we combine Direct Coexistence (DC) simulations with free energy calculations \cite{Adiran2024,adiran2025,Free_en,Jorgensen1988} to investigate how glycine impurities control glyphosate crystallization. Free energy calculations are used to quantify the thermodynamic driving force for crystallization---i.e., relative supersaturation with respect to its solubility limit---estimating the relative glyphosate solvation free energy at different glycine concentrations. Moreover, through DC simulations we model glyphosate crystals in equilibrium with aqueous solutions containing varying concentrations of glycine, providing insights into critical interfacial phenomena while yielding concentration-dependent solubility data. Together, these complementary MD approaches aim to reveal how glycine modulates glyphosate solubility from both thermodynamic and molecular perspectives. Furthermore, our experimental measurements confirm our computational predictions through systematic evaluation of the saturation temperatures and induction times under controlled conditions. These experiments directly quantify glycine's impact on solubility, nucleation kinetics, and growth rates, demonstrating an excellent agreement with computational predicted trends. Our approach, integrating MD modelling and experimental characterization of glyphosate crystallization establishes a relationship between impurity concentration and crystallization behavior in agrochemical production of glyphosate. Specifically, we elucidate the mechanisms through which glycine---an ubiquitous byproduct in glyphosate synthesis---impacts the solubility and disrupts the purification efficiency, providing actionable insights for further process optimization.

\section*{Materials and methods} \label{sec:matnmeth}

\subsection{\label{subsec:meth1}Models and simulation details}

We perform DC simulations using the GROMACS 2023 MD simulation package \cite{gromacs}. For the glyphosate–glycine aqueous solution systems, we use the OpenFF 2.0.0 force field    \cite{Forcefield_1,OpenFF,SMIRNOFF} using a total potential energy function ($U_{total}$) that corresponds to the sum of the bonded and non-bonded interactions between the different atoms. For the non-bonded interactions, a 12-6 Lennard-Jones potential and Coulombic interactions are used. Water is modelled with the same type of potential interactions using the TIP3P water model included in the OpenFF 2.0.0 force field. Cross-interactions are handled via Lorentz-Berthelot mixing rules \cite{Lorentz1881,Berthelot1898,Jorgensen1996} according to the OpenFF 2.0.0. The LINCS algorithm \cite{LINCS} is used to ensure the bond constraints to any hydrogen atom. The molecular topologies used in this simulations were generated with the OpenFF toolkit \cite{OpenFF}. All simulations are performed at constant pressure (p=1 bar) and temperature (T=300 K). Direct Coexistence simulations are performed in the $NpT$ ensemble using an anisotropic Parrinello-Rahman barostat \cite{parrinello1981polymorphic} with a relaxation time of 10 ps and a v-rescale thermostat \cite{Bussi2007} with a relaxation time of 0.1 ps. The time step for the Leap-frog algorithm \cite{berendsen1984leap} is 2 fs. For further details on the force field parameters including the potential cut-offs, and details on the Ewald summation algorithms (PME) for the long-range contribution of the electrostatic interactions please see the Supplementary Material (SM) Section S1.

For the DC simulations, we place a bulk of crystal glyphosate pre-equilibrated on one side of the simulation box, and water (or a glyphosate solution) on the remaining space of the simulation box avoiding intermolecular overlapping between both phases \cite{frenkel2023understanding,espinosa2013fluid,sanchez2024predictions,sanchez2021parasitic,garaizar2024toward}. \textcolor{black}{The exposed crystal orientation is either the (010) or the (001), which we specify in each corresponding section.} We carry out two types of DC simulations: (i) including a perfect crystal of glyphosate in contact with a glyphosate solution; (ii) a glyphosate crystal with random vacancies introduced at the interface in contact with the solution. Once the system reaches equilibrium---normally after 100 nanoseconds---we compute density profiles along the long axis of DC simulation box using the GROMACS tool \emph{gmx density} \cite{gromacs}. Then the solubility ($m_{glyphosate}$) is estimated as:

\begin{equation}\label{equation:solubility}
    m_{glyphosate}=\frac{\rho^{glyphosate}}{\rho^{H_2O}\cdot M_{glyphosate}},
\end{equation}

where $\rho^{glyphosate}$ and $\rho^{H_2O}$ are the mass density of glyphosate and water in the solution respectively, and $M_{glyphosate}$ represents the molar mass of glyphosate.

We additionally perform Free Energy Perturbation (FEP) MD calculations to estimate the solvation free energy of glyphosate in a series of different solutions with varying concentrations of glycine. This methodology provides accurate relative solubilities of small molecule active ingredients in a wide range of solvents and solutions as recently shown in Refs.  \cite{Adiran2024,Free_en,adiran2025}. For these calculations we use the Schr\"{o}dinger’s FEP+ implementation \cite{harder2016opls3e,wang2015accurate} to visualize and setup atomistic simulations with 0.5\%, 1\% and 2\% glycine weight percentages and a single molecule of glyphosate. For these calculations all interactions were parametrized using the OPLS4 force field \cite{roos2019opls4}, a well-established force field widely employed in free energy calculations \cite{Adiran2024}. In these calculations, the solvation free energy is obtained from thermodynamic hamiltonian integration \cite{vega2008determination} through a combination of MD runs and post-processing analysis using the Bennett Acceptance Ratio (BAR) method \cite{tuckerman1992reversible}, which provides an efficient estimator of free energy differences from forward and reverse perturbations. We first equilibrate the different systems by running a ladder of different relaxation simulations as described in the Schr\"{o}dinger's solvation protocol \cite{schrodinger2023-1,adiran2025}. Temperature is kept constant throughout the simulations using a Nose–Hoover thermostat \cite{martyna1992nhc} with a relaxation time of 1 ps. The pressure is kept constant at $p=1$ bar with an MTK barostat \cite{martyna1994mtk} with a relaxation time of 2 ps. We sample 20 different replicas of each glycine weight percentage, allowing us to sample different local environments to accurately represent solution's bulk properties. Further details regarding FEP+ calculations are provided in Section S2 of the SM.

The link between the solvation free energy and solubility (described in Equation \ref{equation:free-en}) allows an indirect estimation of the relative solubility under different solution conditions:

\begin{equation}\label{equation:free-en}
    \ln\!\left( \frac{m_2}{m_1} \right) = \beta \left( \Delta G_1^{\infty \:\text{solv}} - \Delta G_2^{\infty\:\text{solv}} \right),
\end{equation}

where $m_1$ and $m_2$ represent the solubilities of a given solute in different solution conditions, $\beta$ is the reciprocal of $k_BT$, where $T$ is the temperature in Kelvin and $k_B$ is the Boltzmann constant, and $\Delta G_n^{\infty\:\text{solv}}$ is the solvation free energy of the compound in a given solvent $n$ at infinite dilution.

\subsection{\label{subsec:mat1}Materials and experimental Methods}

Glyphosate has been provided by the Bayer Crop Science department, with HPLC grade \>95\% purity.
We measure the solubility and detection time using a Crystal16 (Technobis Crystallization Systems, 2021). Here the detection time is defined as the time elapsed between reaching the supersaturation point and the first detection of crystals. Solubilities are measured according to the following protocol: in a vial selected amounts of glyphosate, glycine, phosphoric acid, and water ware weighed and allowed to equilibrate at 293 K for three hours. The temperature is then gradually increased at a rate of 1 K/min up to 333 K, and the point at which the solution becomes clear (the clear point) was interpreted as the saturation temperature. For the water-glyphosate system, the solubility measurements obtained via the clear point method were also confirmed by equilibrating a suspension at 304 K for 24 hours, and measuring the concentration of the liquid phase via HPLC.

We determine the detection time with Crystal16, which is equipped with a laser to detect variations of light transmittance in the solution being monitored. This method is well-established experimentally to infer nucleation kinetics and it has been extensively discussed in literature. Further details on the method, assumptions, and limitations can be found in Kadam \textit{et al.} and Maggioni \textit{et al.} \cite{maggioni2017,maggioni2017b,kadam2012}. Specifically, each cycle consists of a rapid heating from 293 K to 333 K (30 min), a hold time at 333 K (60 min), a rapid cooling back to 293 K (1 min), and a subsequent hold time at 293 K (360 min). Then, the detection time is evaluated from the moment 293 K was attained after the rapid cooling. Because crystallization involves both nucleation and growth, which cannot be separately resolved with this technique, we adopt a simplified analysis. For each condition~$j$ with detection times~$I_j$, we define the growth time as

\begin{equation}
  t_{\text{growth},j} = \min_{i \in I_j} \left[ t_{ji} \right], 
\end{equation}
i.e., the first observed detection time, assuming nucleation occurs at $t=0$. The nucleation time for each experiment~$i$ is then calculated as 

\begin{equation}
    t_{n,ji} = t_{ji} - t_{\text{growth},j},
\end{equation}
assuming that growth time is constant under identical conditions. In this manner, nucleation and growth contributions can be distinguished, and their sum corresponds to the overall detection time. Since our aim is to capture qualitative trends rather than exact nucleation kinetics, this approximation is appropriate for the present analysis.

\section*{Results and discussion\label{sec: Results}}
\addcontentsline{toc}{section}{Results and discussion} % optional
\stepcounter{section}

\subsection{Simulations capture crystal density but underestimate glyphosate solubility}

\begin{figure*}[hbt!]
	\centering
	\includegraphics[width=0.99\linewidth]{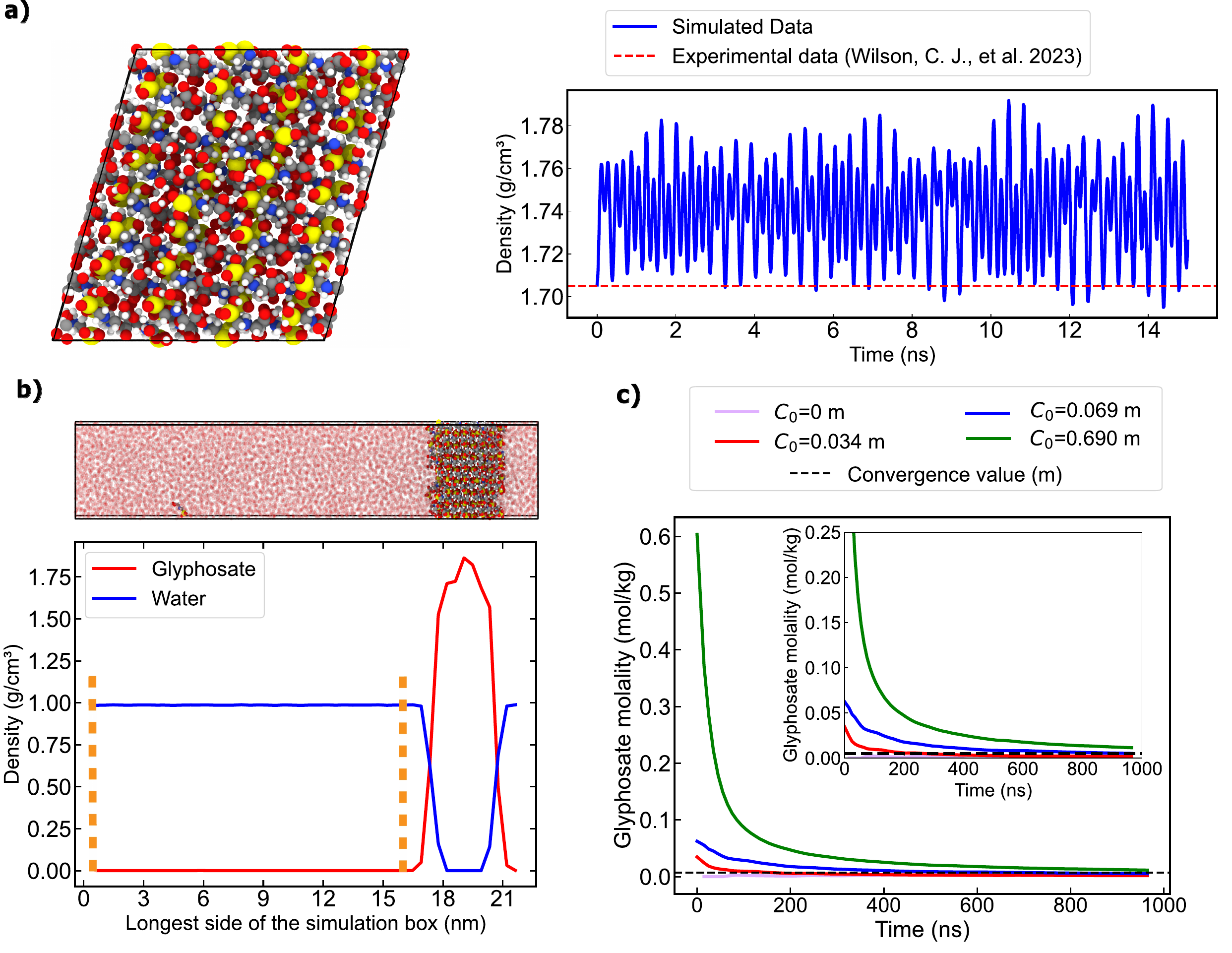}
	\caption{a)~Left: Snapshot of the glyphosate bulk crystal structure \cite{knuuttila1979crystal} from our simulations. Right: Time-evolution of the crystal density from $NpT$ simulations at 300K compared to the experimentally determined value from Ref. \cite{wilson2023discerning} (dashed red horizontal line). b)~Top: Representative snapshot from a DC simulation where water molecules have been rendered semi-transparent to better visualize the glyphosate molecules in the bulk. Bottom: Density profile along the longest box direction of one of our systems ($C_0 = 0.069 \text{m}$) without vacancies upon reaching equilibrium. The density profiles have been separated into different species as indicated in the legend. The region that we define as the \textit{bulk} solution is in between the two dashed orange vertical lines. c)~Time-evolution of the glyphosate average concentration in the solution for different systems, each with a different initial glyphosate concentration in the bulk liquid phase. The values were obtained by calculating the converged average value of the concentration at all times $<n$ time. The average value among all the different trajectories (i.e., solubility limit, $m$) is indicated with a dashed horizontal line. The inset shows a zoomed-in view of the Y-axis.}
 \label{fig:fig1}
\end{figure*}

Initially, we assess the force field's ability to reproduce well-established properties of glyphosate, such as the crystal density, by preparing a $4 \times 4 \times 4$ replication of the crystal unit cell \cite{knuuttila1979crystal} (Fig.~\ref{fig:fig1}a, left) at 300~K and 1~bar in the $NpT$ ensemble. We then study how the bulk crystal density fluctuates throughout a simulation and compare it to the experimental data obtained by Wilson \textit{et al.} \cite{wilson2023discerning}, obtaining values in reasonable agreement (Fig.~\ref{fig:fig1}a, right panel). \textcolor{black}{The small overestimation of the simulated density ($\sim$2\%) is within the typical accuracy of a general-purpose fixed-charge force field \cite{boothroyd2023development}, which has not been explicitly parametrized for a single molecule (e.g., glyphosate) crystal properties.} Moreover, we determine that the crystal structure is well conserved throughout the MD simulation, preserving the crystal’s structural integrity and density over time.

Once we confirm that the crystal is accurately modelled and that correctly maintains its crystal structure, we continue by preparing DC initial configurations \textcolor{black}{exposing the (010) crystal orientation}, where the crystal structure is in direct contact with different aqueous solutions, which contain varying concentrations of glyphosate in the solution. We corroborate that the initial glyphosate concentration in the solution does not affect the obtained solubility determination by performing multiple simulations with different initial concentrations of glyphosate. This is because in the NpT ensemble, the system spontaneously evolves towards equilibrium conditions, in which the crystal coexists with the solution at the saturation concentration \cite{espinosa2016calculation}. The prepared DC configurations contained 8000 water molecules and the necessary varying glyphosate replicas in the solution ranging from 0 mol/kg to 0.690 mol/kg (Further details of system sizes can be found in SM section SIX).

As discussed in the \hyperref[sec:matnmeth]{Methods Section}, we use Eq. \ref{equation:solubility} to determine the solubility from the density profile along the long axis of the DC simulations, as illustrated in  Fig.~\ref{fig:fig1}b. Through orange dashed vertical lines, we bound the liquid phase from where we obtain the parameters that enter into Eq. \ref{equation:solubility} to determine the glyphosate concentration in the solution ($m$). In Fig.~\ref{fig:fig1}c, we show the time evolution of the glyphosate concentration (in molality units) within the solution. While all the different simulations significantly differ in the initial glyphosate concentration, all trajectories converge to similar values within the uncertainty upon $\sim$600 ns. Importantly, we note that although stochasticity plays a role in the time needed for a given initial configuration to converge in the equilibrium state, such effect becomes more pronounced when the initial solution concentration largely differs to the solubility limit \cite{espinosa2016calculation} as for the simulation starting from $C_0$ = 0.690 m (green curve in Fig. \ref{fig:fig1}c).

Furthermore, we perform DC simulations\textcolor{black}{, also exposing the (010) orientation,} to examine how the presence of vacancies affect the kinetics to reach the equilibrium concentration of glyphosate in the solution. To build these systems, we randomly remove 10 glyphosate molecules from the outer crystal plane in contact with the solution, artificially creating defects on the surface. The same initial concentrations of the glyphosate solution depicted in Fig. \ref{fig:fig1}c are then used to determine the solubility from the density profiles once the systems reach equilibrium (i.e., the concentration of glyphosate in the solution shows no drift as a function of time). The obtained average results from these two types of simulations using multiple trajectories with different initial concentrations are summarized in Table \ref{tab:sol} (further details on these calculations are shown in Figures S1, S2 and Table S1). Based on these results, we do not find significant differences neither in the kinetics nor in the equilibrium value of the DC simulations in presence \textit{vs.} absence of interfacial defects. Nevertheless, we note that in most of our defect-free simulations, the glyphosate molecules incorporated from solution into the crystal do not adopt the correct bulk lattice structure.

\begin{table*}[!ht]
\centering
\caption{\textcolor{black}{Comparison of the experimental density\cite{wilson2023discerning} and solubility\cite{gly_sol} predictions, along with computational calculations (using DC simulations with a crystal phase in absence versus presence of vacancies at the interface).}}
\label{tab:sol}
\begin{tabular}{|c|c|c|}
\hline
                           & Crystal Density (g/$cm^3$) & Solubility (mol/kg)      \\
\hline
Experimental data           & $1.705$          & $(62.1 \pm 0.1)\cdot10^{-3}$ \\
\hline
Simulated without vacancies & $1.74 \pm 0.04$ & $(4 \pm 3)\cdot10^{-3}$        \\
\hline
Simualted with vacancies    & -                & $(4 \pm 3)\cdot10^{-3}$       \\
\hline
\end{tabular}
\end{table*}

We then examine how the glyphosate solubility obtained via DC simulations compares with experimentally reported values. Table~\ref{tab:sol} summarizes both the experimental results~\cite{gly_sol,wilson2023discerning} and our computational predictions for crystal density (in $g/cm^3$) and solubility (reported in millimolality for consistency with the reference data). The OpenFF force field reproduces the crystalline density within 2\% of the experimental value but significantly underestimates solubility, by roughly an order of magnitude. This discrepancy is consistent with the general parametrization strategy of OpenFF, which does not explicitly target organic molecule solubilities \cite{wilson2023discerning}. Indeed, even for simpler aqueous systems such as NaCl solutions, state-of-the-art force fields often underestimate solubilities, with deviations ranging from 10\% up to an order of magnitude~\cite{espinosa2016calculation,benavides2016consensus,sanchez2023direct,benavides2017potential}. Importantly, the convergence of simulations initiated at different glyphosate concentrations toward similar equilibrium solubility values reinforces that OpenFF systematically underestimates glyphosate solubility.

\subsection{Calculation of glyphosate solubility using DC simulations exposing different crystal faces} 

\begin{figure*}[hbt!]
	\centering
	\includegraphics[width=.9\linewidth]{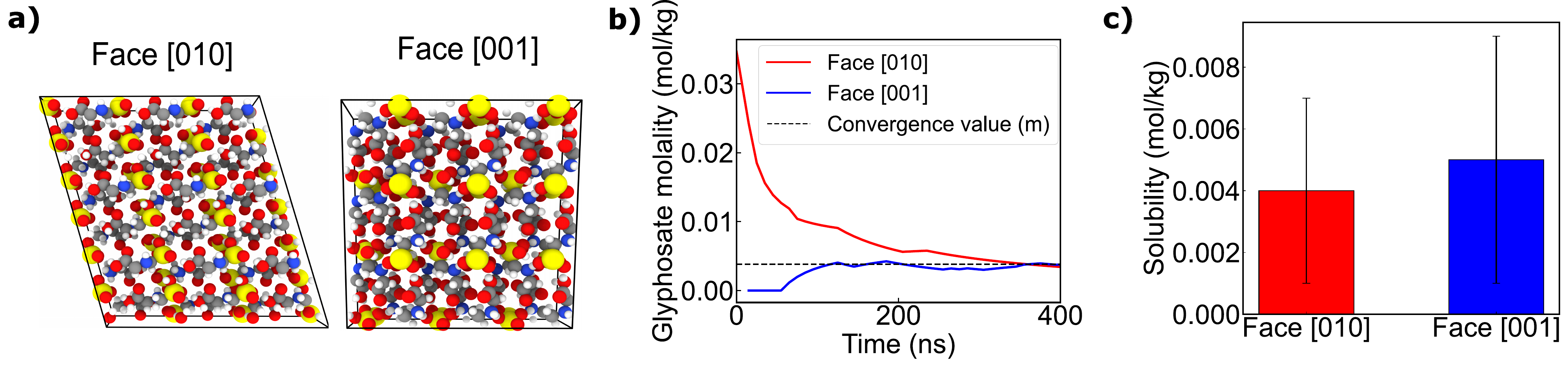}
	\caption{a)~Snapshots of the different crystallographic planes studied in a 4 x 4 x 4 unit cell. b)~Convergence of the molality of two independent DC simulations with different crystal orientations and different initial glyphosate concentrations as a function of time. The values were obtained by calculating the arithmetic accumulated average value of the solubility at all times $<n$ time. The average value of both trajectories is depicted by an horizontal dashed line. c)~Bar plot of the obtained solubility of the two different crystallographic faces. The whisker represents the uncertainty bounds.}
\label{fig2}	
\end{figure*}

As previously discussed, glyphosate crystallizes from aqueous solution in a pH-dependent zwitterionic form, adopting a monoclinic unit cell. This structure belongs to the $P2_{1}/c1$ space group in the Hermann–Mauguin notation. As shown in Fig.~\ref{fig2}a, crystal faces with different Miller indices exhibit distinct characteristics, such as the number of molecules directly exposed to the solvent or their interfacial free energy with the solution. In Section A, we have focused on systems where the (010) crystal face of glyphosate was in contact with the aqueous phase. In the present section, we investigate whether exposing a different crystal face, as illustrated in Fig.~\ref{fig2}a, may affect the determination of solubility. From a thermodynamic standpoint, the solubility limit of a crystalline solid in solution is independent of which crystal face is exposed. At equilibrium, the chemical potentials of the solid and dissolved species are equal, and solubility is therefore a bulk property of the coexisting phases rather than a surface-dependent quantity~\cite{frenkel,espinosa2016calculation}. Nevertheless, different faces present distinct interfacial free energies and growth rates \cite{montero2019ice}, which can give rise to anisotropic growth and dissolution rates \cite{Freitas2020,sanchez2021parasitic}. In practice, such differences may influence kinetics, even though they should not alter the equilibrium solubility itself. Moreover, comparing solubility values obtained from simulations using different crystal orientations provides an important internal consistency check. If calculations with distinct faces converge to the same solubility, this rules out the possibility that finite-size effects or kinetic trapping are biasing the results, since it is highly unlikely that two independent orientations would be affected by the same artifact in an identical manner. Thus, studying multiple orientations not only probes potential kinetic effects but also strengthens confidence in the robustness of the solubility determination.

To investigate potential kinetic trapping and finite-size effects in our systems, we consider an additional crystal orientation, (001), and construct two new systems for DC simulations. In both cases, the number of water molecules was fixed at 8000, and the required number of glyphosate molecules in the solution was added to achieve initial concentrations of $0\:\text{m}$ and $0.69\:\text{m}$. As shown in Fig.~\ref{fig2}b, systems differing in both the exposed crystal plane and the initial glyphosate concentration converge to the same solubility value, indicated by the dashed line. Statistical analysis of the average solubilities for both orientations (Fig.~\ref{fig2}c) shows that the results are equivalent within the associated uncertainties. In summary, the crystal face in contact with the aqueous solution does not affect the solubility of glyphosate within the uncertainty, consistent with previous observations in other systems using DC simulations~\cite{espinosa2016calculation}. This agreement between orientations indicates that our simulations are not kinetically trapped and that finite-size effects are negligible.

\subsection{Glycine impurities increase glyphosate solubility}

\begin{figure*}[hbt!]
	\centering
	\includegraphics[width=\linewidth]{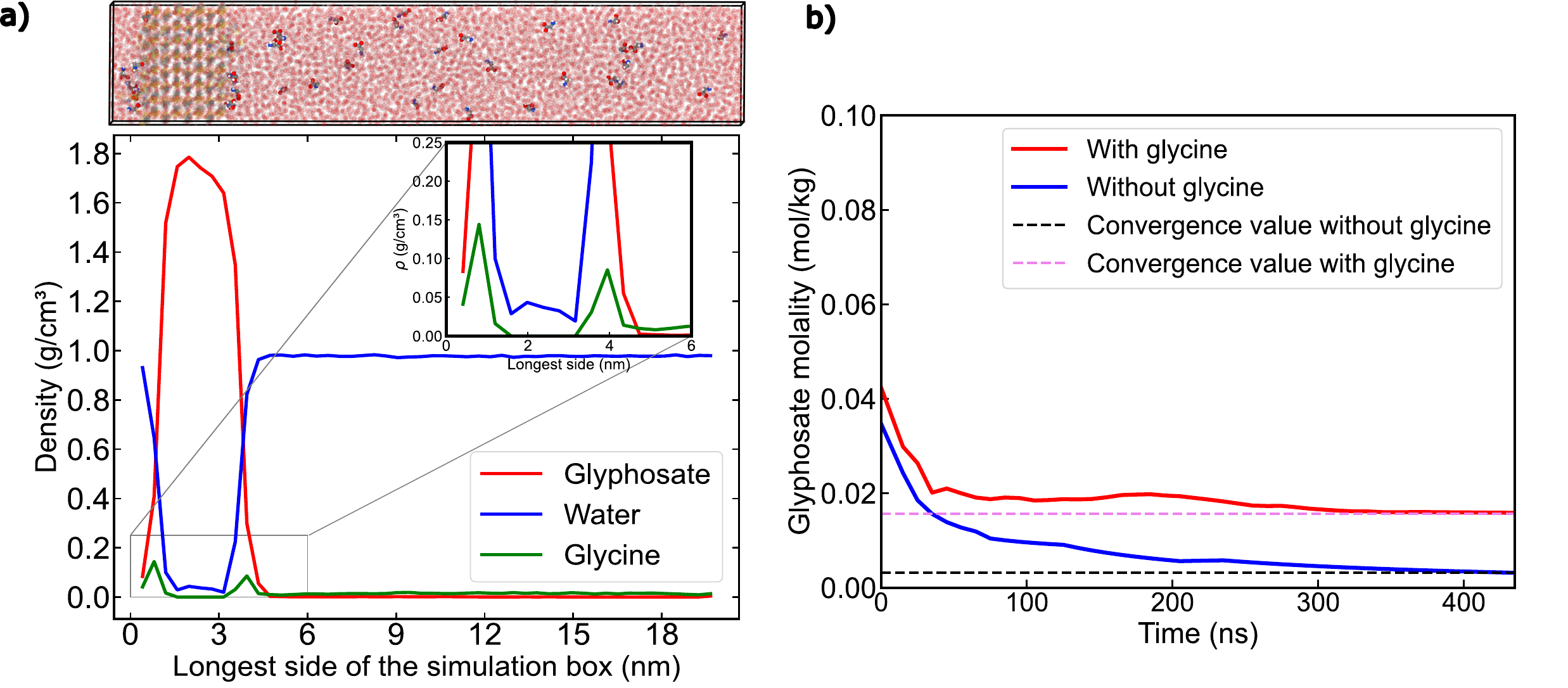}
	\caption{a)~Top: Snapshot from a DC simulation where water and glyphosate molecules have been rendered semi-transparent to better visualize the glycine molecules in the bulk and at the crystal surface. Bottom: Density profile of a DC simulation along the longest box direction, separated into its different components as indicated in the legend. The inset presents a zoom-in insight of the density profile showing the glycine density peaks at the crystal surface. b)~Glyphosate concentration \textit{vs.} time for two different systems, one containing 2\%wt glycine (shown in red), and another in absence of glycine (shown in blue). The values were obtained by calculating the accumulated average value of the glyphosate concentration at times $<n$ time. The convergence value (solubility limit, $m$) is indicated with a dashed horizontal line (shown in violet and black in presence \textit{vs.} absence of glycine, respectively). 
    }
	\label{fig3}
\end{figure*}

Real systems are more complex than just water and glyphosate, as they often contain glycine, a ubiquitous impurity introduced during glyphosate synthesis. At the pH where glyphosate crystallizes, glycine exists mainly in two forms: as a zwitterionic species and a negatively charged species, in a 1:4 ratio~\cite{yu2002glycine}. To examine the effect of this impurity, we construct four systems with varying glycine concentrations while maintaining a fixed number of water and glyphosate molecules in the solution. In addition, we consider both crystal orientations previously described in Sections A and B of the Results. As before, we analyze the DC simulations and obtain the density profiles of the new systems, shown in Fig.~\ref{fig3}a. Glycine is dynamically adsorbed onto the crystal surface through a continuous adsorption–desorption process, as evidenced by the two maxima highlighted in the inset of Fig.~\ref{fig3}a. This behavior is observed for both glycine forms (zwitterionic and negatively charged; see Fig.~S4). The snapshot in Fig.~\ref{fig3}a further reveals that, instead of aggregating into a solid-like structure on the glyphosate crystal, glycine forms a partial coating layer at the interface, characteristic of transient adsorption. \textcolor{black}{We further characterize this behaviour by measuring the surface concentration $\Gamma = N_{\mathrm{Glycine}}/A$ for the (010) and (001) planes at two glycine concentrations (0.5 and 2~wt\%). Our surface concentration measurements through DC simulations are summarized in Table~\ref{tab_adsorption}. Based on our calculations, for a given concentration, the different exposed crystal planes exhibit similar adsorption capacities, as the values of $\Gamma$ agree within the associated uncertainties. Nevertheless, higher glycine concentration leads to a greater surface concentration, a behaviour that is known to further hinder crystal growth in similar systems \cite{salvalaglio2012uncovering,tanaka2022crystal}. This prediction will be later confirmed in results from Section D, where we experimentally determine the crystal growth rate as a function of different glycine concentrations.}
\\

\begin{table}[]
    \centering
    \begin{tabular}{c|c|c}
         \multirow{2}{*}{Plane} &  \multicolumn{2}{c}{$\Gamma$ (molecules/nm$^2$)} \\ \cline{2-3}
         & 0.5 \% wt Glycine & 2 \% wt Glycine \\ \hline
         (010) & 0.11(10) & 0.36(10) \\ \hline
         (001) & 0.18(10) & 0.43(10)
    \end{tabular}
    \caption{\color{black}Surface concentration of glycine ($\Gamma$) for two different crystal orientations of glyphosate measured at two glycine concentrations using DC simulations.}
    \label{tab_adsorption}
\end{table}

In Fig.~\ref{fig3}b, we display the time-evolution of glyphosate concentration converging to its solubility value in the presence (2~wt\% glycine) and absence of glycine. The presence of glycine increases the solubility by approximately an order of magnitude. Importantly, this enhancement in solubility cannot be explained solely by the transient surface coating observed in Fig.~\ref{fig3}a hindering glyphosate attachment into the surface. We hypothesize that glycine adsorption may result from its concentration exceeding the solubility limit, leading to heterogeneous nucleation at the glyphosate–water interface. However, a detailed investigation of glycine solubility is beyond the scope of this study.

\begin{figure*}[hbt!]
	\centering
	\includegraphics[width=\linewidth]{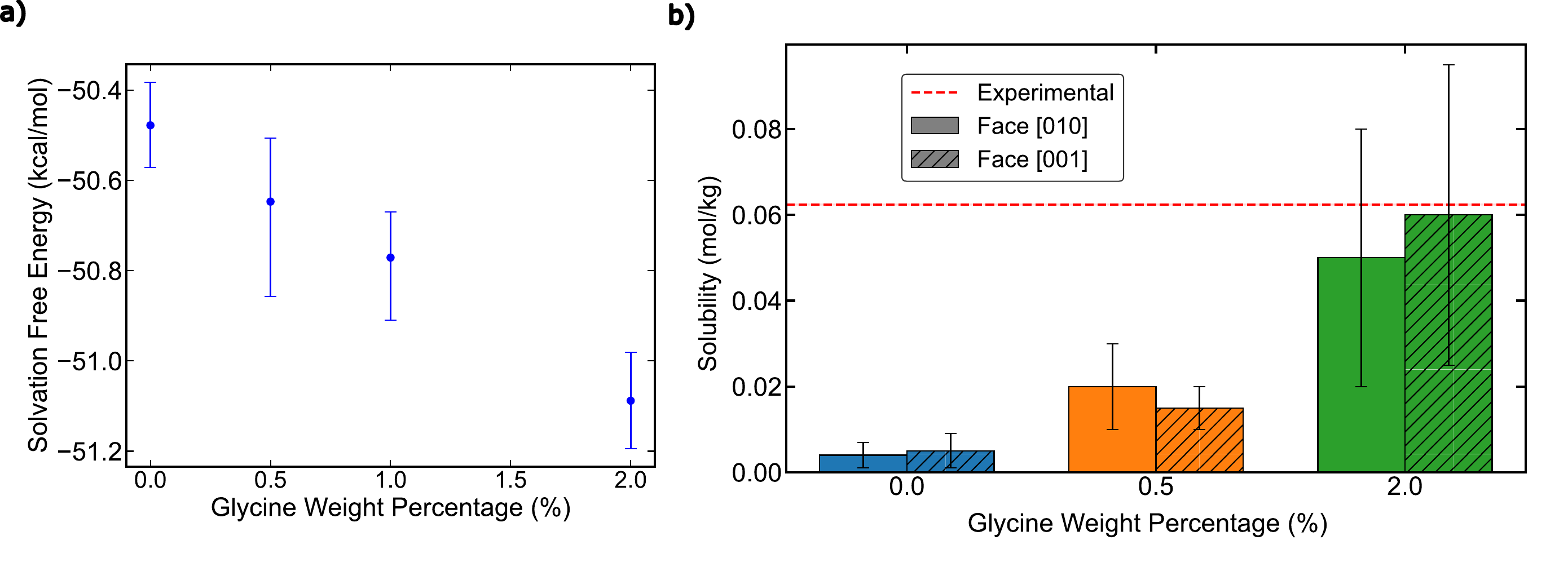}
	\caption{a)~Solvation free energy as a function of glycine weight percentage from free energy calculations. We observe that increasing glycine content leads to a further decrease in solvation free energy. b)~Bar plot of the obtained solubility in the different studied systems, \textcolor{black}{with different glycine concentrations. The horizontal dashed line depicts the experimental solubility value reported in absence of impurities \cite{gly_sol}.}  
    }
	\label{fig4}
\end{figure*}

As commented in Section B, solubility is determined by the equality of the chemical potentials of the crystal and dissolved molecules in the solution~\cite{frenkel}. The presence of glycine at the interface does not alter the chemical potential of glyphosate in either phase and, therefore, should not directly affect solubility. Instead, glycine adhesion at the interface is expected to influence only the dynamic properties of the crystal, such as growth and exchange rates. To further investigate the origin of the solubility change, we examine how glycine affects the solvation free energy of glyphosate in water. Solvation free energy is defined as the change in the Gibbs free energy when a solute is transferred from vacuum to a given solvent/solution at constant temperature and pressure. It quantifies the favorability of solute–solvent interactions. We calculate glyphosate solvation free energies using the FEP+ method described in Section~\hyperref[subsec:meth1]{Models and simulation details}. The systems spanned glycine concentrations from 0~wt\% to 2~wt\%, and each contained 15,000 water molecules, one glyphosate molecule, and the corresponding number of glycine molecules required to reach the target concentration. The results, shown in Fig.~\ref{fig4}a, reveal a clear trend: the presence of glycine decreases the solvation free energy of glyphosate.

Therefore, the presence of glycine makes it more favorable for glyphosate to remain in solution compared to the impurity-free system. Moreover, this analysis reveals an inverse correlation between glycine concentration and solvation free energy: as glycine concentration increases, the solvation free energy decreases. 
%The full trajectories of all systems studied here are provided in Section~S6 of the SM. 
This result explains the increase in solubility observed in Fig.~\ref{fig3}b when glycine is added. In this case, the presence of impurities directly alters the chemical potential of glyphosate in solution. The change in solvation free energy is therefore linked to the solubility of the system through Eq.~\ref{equation:free-en}. 
\textcolor{black}{This mechanism is also corroborated in Fig.~\ref{fig4}b, where we compare the solubility obtained with increasing glycine concentrations. We observe that the presence of glycine increases glyphosate solubility monotonically by approximately an order of magnitude within the studied concentration range. Moreover, we also observe that, as expected, the solubility is not affected by the crystal orientation exposed to the solution, as already established in Results Section B. Additionally, from our DC simulations, we computed the radial distribution functions between the different heavy atoms of glyphosate and glycine in order to determine which specific interactions are responsible of their intermolecular association that drives the solvation of glyphosate and adsorption of glycine to the crystal. These distributions are shown in Figure S7, and reveal that electrostatic interactions are crucial, being that between glyphosate's oxygen and glycine's nitrogen the most relevant one increasing both the solvation free energy (in absolute value) and the solubility as glycine concentration increases (see SM Figure S7).}

\textcolor{black}{Further comparison} of our simulations with experimental data suggests that glycine, frequently present in glyphosate samples~\cite{zhou2012study}, may significantly influence solubility measurements. Using Eq.~\ref{equation:free-en} and the 2~wt\% glycine data, we estimate the solvation free energy corresponding to the experimentally reported solubility ($62.4 \pm 0.1$~mmol/kg). Assuming a linear correlation between glycine concentration and solvation free energy (Fig.~\ref{fig4}a), we predict an effective glycine content of $2.7 \pm 0.5$~wt\% in the experimental system assuming that the OpenFF would predict the correct solubility for glyphosate. Although the actual glycine content of the experimental samples was not reported~\cite{gly_sol}, this estimation provides a plausible explanation for the observed discrepancy in solubility---beyond force field deficiencies---and further strengthens the value of our computational framework for characterizing and predicting glyphosate phase behavior.

In summary, our results show that glycine hinders glyphosate crystallization through two distinct mechanisms. First, glycine dynamically adsorbs at the crystal--solution interface, coating the surface and potentially altering growth and exchange dynamics \cite{sanchez2021parasitic, mutaftschiev2013adsorption,sangwal1996effects}. Second, glycine lowers the solvation free energy of glyphosate, thereby making dissolution thermodynamically more favorable. Overall, glycine promotes glyphosate solubility in water by primarily modifying the solution thermodynamics.

\subsection{Experimental validation of the simulation predictions \label{subsec: Result4}}

\begin{figure*}[hbt!]
	\centering
	\includegraphics[width=\linewidth]{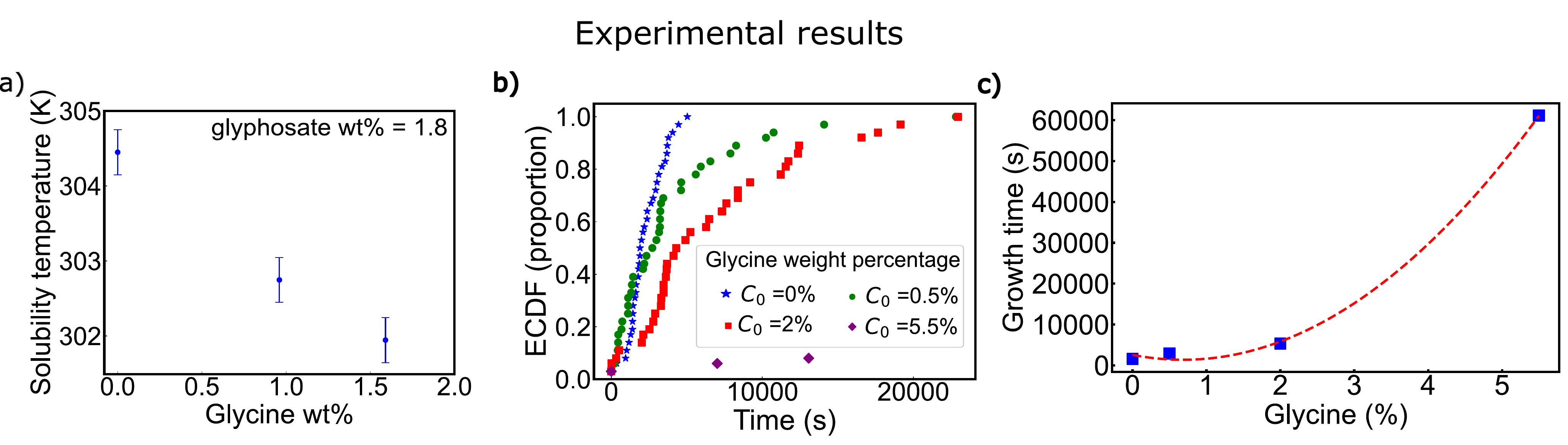}
	\caption{a) Solubility temperatures (measured as clear points) as a function of glycine percentage in weight. Experiments are performed at constant concentration of glyphosate in acidic water for increasing concentrations of glycine. Our data refer to the mean values of triplicates, with the error bars indicating the associated standard deviation. b) Empirical cumulative distribution functions (ECDFs) of crystallization times obtained from the measured detection times at constant glyphosate supersaturation ($S=2$), and varying concentrations of glycine as indicated in the legend. The experiments are performed at constant temperature, $T=293K$. c) Growth times obtained from the detection time analysis against the glycine weight percentage in the solution. A second-order fit, shown as a dashed red line, is included as a guide for the eye. The experimental uncertainty of these measurements is of the order of the symbol size.}
	\label{fig5}
\end{figure*}

To validate our simulation predictions, we conduct targeted experiments addressing two main objectives: (i) to confirm the enhancement of glyphosate solubility in the presence of glycine, and (ii) to assess whether crystal formation kinetics are also slowed down, based on our hypothesis that glycine adsorbed at the crystal surface can interfere with crystal growth (Figure~\ref{fig3}). For this purpose, the clear point temperature (equivalent to the saturation point) is determined in triplicate at constant glyphosate molality while varying the glycine concentration, following the procedure described in Section~B of the \hyperref[subsec:mat1]{Materials and Methods}. Figure~\ref{fig5}a shows the variation in solubility temperature as a function of glycine concentration. The results indicate that glycine lowers the solubility temperature of glyphosate, thereby enhancing its solubility. Consequently, at a constant temperature, a higher glyphosate concentration is required to reach saturation, in full agreement with our simulation predictions from Figure~\ref{fig3}.

Furthermore, to investigate the effect of glycine on the kinetics of glyphosate crystallization, we employ an indirect measurement approach based on detection times (see Section~B of the Materials and Methods for further details on the determination of this quantity). Regardless of the specific mechanism inhibited by glycine---primary nucleation or crystal growth---we hypothesize that the detection time distribution would shift towards longer times and may also broaden. To test this, we prepare systems with varying glycine concentrations and adjusted the corresponding glyphosate concentration to maintain a constant supersaturation of $S=2$ at 293~K. The detection time is composed of two contributions: the nucleation time $t_N$ and the growth time $t_G$. The nucleation time corresponds to the interval between reaching the saturation point and the formation of the first nucleus, whereas the growth time is the interval between the nucleus formation and the detection of the crystals. While $t_N$ is an intrinsic hardly attainable property of the system, $t_G$ depends on the monitoring technique employed. Further methodological details can be found in Kadam \textit{et al.} and Maggioni \textit{et al.} \cite{maggioni2017,maggioni2017b,kadam2012}. By applying the detection time analysis described in the Materials and Methods section, we evaluate the distributions of nucleation and growth times from the overall detection time distributions. Figure~\ref{fig5}B shows the empirical cumulative distribution functions (ECDFs) of nucleation events at three different glycine concentrations. The nucleation time distributions broadened and their medians shifted towards larger values, indicating that nucleation became slower (i.e., less likely to occur within a fixed time), consistent with a reduced nucleation rate. Moreover, Figure~\ref{fig5}C also presents the calculated growth times as a function of glycine concentration. Here, we observe that the growth time increases with glycine concentration, suggesting that the transformation of nuclei into observable crystals is also slower. Since the growth time primarily reflects the growth rate, these results indicate that the crystal growth rate decreases as the glycine concentration increases.

Taken together, our experimental observations show remarkable consistency with the results of our molecular dynamics simulations. This agreement is significant, as it not only validates the predictive capability of our simulation framework but also provides a molecular-level explanation for the observed macroscopic phenomena. Specifically, our simulations confirm that glycine plays a dual mechanistic role in modulating glyphosate crystallization. On the one hand, glycine alters the thermodynamic properties of the solution, enhancing solubility and thereby reducing the driving force for nucleation, as reflected in the broadening and temporal shift of the nucleation time distributions. On the other hand, the simulations reveal that glycine molecules preferentially co-localize at the surface of nascent glyphosate crystals. This surface association offers a plausible mechanistic explanation for the experimentally observed increase in growth time, as the adsorbed glycine molecules are likely to hinder the incorporation of glyphosate molecules into the growing crystal lattice. The outcome is a reduction in the effective growth rate, consistent with the experimental trends. Thus, the synergy between experiments and simulations enables us to propose a comprehensive mechanistic picture. The combined effects highlight the multifaceted role of impurities in crystallization processes and emphasize the importance of integrating molecular-level simulations with experimental validation. Beyond confirming the robustness of our findings, this approach provides a powerful framework for the rational design of crystallization modifiers in complex multicomponent systems.

\section*{\label{sec:conclusion}Conclusions}

In this work, we have investigated the molecular mechanisms by which glycine---a ubiquitous synthesis impurity---modulates the thermodynamics and kinetics of glyphosate crystallization in aqueous solution. By combining Direct Coexistence molecular dynamics simulations, free energy calculations, and experimental detection times of crystal growth, we show that glycine acts not as an inert byproduct but as an active crystallization inhibitor operating through dual pathways. Direct coexistence simulations reveal dynamic glycine adsorption at crystal–solution interfaces, forming transient partial layers that contribute to blocking glyphosate incorporation sites and hinder growth attachment. In parallel, free energy calculations demonstrate that glycine decreases glyphosate’s solvation free energy, thereby enhancing equilibrium solubility---as also demonstrated through Direct Coexistence simulations---and reducing supersaturation, the fundamental thermodynamic driving force for nucleation and growth. Together, these synergistic mechanisms establish glycine as both a kinetic inhibitor and a thermodynamic modulator of glyphosate crystallization.

Our simulations further show that glyphosate solubility is invariant across different crystallographic faces and insensitive to surface defects, confirming that glycine’s effects originate from bulk solution thermodynamics rather than interfacial artifacts. Experimental measurements quantitatively validate these predictions: increasing glycine concentration systematically enhances solubility while hindering both nucleation and growth rates. This convergence between computation and experiment provides a robust mechanistic validation. Overall, this work demonstrates the power of molecular simulations to disentangle the complex interplay between interfacial phenomena and solution thermodynamics in impurity-mediated crystallization. The mechanistic insights presented here---particularly the dual kinetic–thermodynamic inhibition pathway---offer molecular-level design principles for optimizing industrial crystallization processes and predicting impurity effects in pharmaceutical and agrochemical systems. More broadly, our framework establishes a quantitative methodology for understanding how synthetic byproducts influence crystallization in complex aqueous environments.

\section*{Acknowledgements}
A.C. acknowledges funding from  ART. 60 LOSU (259-2024) RESEARCH COLLABORATION AGREEMENT BAYER AG - UNVERSITY and  ART. 60 LOSU (273-2024) RESEARCH COLLABORATION AGREEMENT BAYER AG - UNVERSITY. I.~S.-B. acknowledges funding from  the UK Research and Innovation (UKRI) Engineering and Physical Sciences Research Council (EPSRC) under the UK Government’s guarantee scheme (EP/Z002028/1), following successful evaluation by the ERC (Consolidator Grant awarded to R.C.G.) under the European Union’s Horizon Europe research and innovation programme.  J.~R.~E. acknowledges funding from Emmanuel College, the University of Cambridge, and the Ramon y Cajal fellowship (RYC2021-030937-I). J.~R.~E. also acknowledge the Spanish scientific plan and committee for research; project reference PID2022-136919NA-C33. This work has been performed using resources provided by the Cambridge Tier-2 system operated by the University of Cambridge Research Computing Service (http://www.hpc.cam.ac.uk) funded by EPSRC Tier-2 capital grant EP/P020259/1-CS170. This work has also been performed using resources provided by Archer2 (https://www.archer2.ac.uk/) funded by EPSRC Tier-2 capital grant EP/P020259/e829. The authors also thankfully acknowledge RES computational resources provided by Mare Nostrum 5 through the activities 2024-3-0001 and 2025-1-0009.
G.M.M. thanks Joerg Brockob at Bayer AG for performing the experiments reported in this work.

\section*{Data and code availability}
The data that supports the findings of this study are available within the article and its Supplementary Material. Topology files, mdp files and some initial configurations are available in this \href{https://github.com/Alex-castro-quim/Glyphosate_crystallization}{GitHub link for the repository} ( \href{https://github.com/Alex-castro-quim/Glyphosate_crystallization}{https://github.com/Alex-castro-quim/Glyphosate\_crystallization} ).

\bibliographystyle{ieeetr}

%\bibliography{bibliography}

\clearpage

% \counterwithin{equation}{section}
% \counterwithin{table}{section}
\setcounter{equation}{0}
\setcounter{table}{0}
\setcounter{section}{0}
\setcounter{figure}{0}
\renewcommand\thesection{S\Roman{section}}   
\renewcommand\thefigure{S\arabic{figure}} 
\renewcommand\thetable{S\arabic{table}} 
\renewcommand\theequation{S\arabic{equation}}    
\onecolumngrid
\setcounter{page}{1}
\begin{center}
    {\large \textbf{Supplementary Material: Understanding How Synthetic Impurities Affect Glyphosate Solubility and Crystal Growth Using Free Energy Calculations and Molecular Dynamics Simulations} \\ \par} \vspace{0.3cm}
    Alejandro Castro$^{1,+}$, Ignacio Sanchez-Burgos$^{2,+}$, Nuria H. Espejo$^{1,3}$, Adiran Garaizar$^{3,*}$, Giovanni Maria Maggioni$^{4,*}$, Jorge R. Espinosa$^{1,2,5*}$ \\ \vspace{0.15cm}
    
    $[1]$ Department of Physical Chemistry, Universidad Complutense de Madrid, Av. Complutense s/n, Madrid 28040, Spain \\
    $[2]$ Yusuf Hamied Department of Chemistry, University of Cambridge, Lensfield Road, Cambridge CB2 1EW, UK \\ 
    $[3]$ Data Science, Bayer AG, Alfred-Nobel-Straße 50, 40789 Monheim am Rhein, Germany \\
    $[4]$ Crop Protection Innovation - Bayer AG, Kaiser-Wilhelm-Allee 1, 51373 Leverkusen, Germany \\
    $[5]$ Multidisciplinary Institute, Complutense University of Madrid, Paseo Juan XXIII, 1, Madrid 28040, Spain \\
 
* = To whom correspondence should be sent.
email: adiran.garaizarsuarez@bayer.com , john.maggioni@bayer.com , jorgerene@ucm.es
\end{center}
\thispagestyle{empty}

\section{The OpenFF model \label{supp_model}}

% \subsubsection*{\textbf{Mpipi-Recharged model for disordered, globular, and multi-domain proteins}.}

We use the force field created by OpenFF toolkit\cite{openff_initiativeAA,Forcefield_1AA}.
This initiative uses the Sage 2.0.0 force field format, defining molecular mechanic force fields using standard classical mechanic energy terms with a special parameter assignment system, a SMIRKS-based chemical perception of atom types.
SMIRKS is a chemical substructure query language closely related to SMILES, instead of assigning atom types, OpenFF uses SMIRKS patterns to directly match chemical environments, so each parameter in the force field us attached to a SMIRKS pattern, not a defined atom type.\cite{bannan_chemperAA}
This effectively reduces the number of parameters needed and makes force fields eased to extend and maintain.
The potential energy is computed as the sum of pairwise bonded ($U_{\textrm{bonded}}$) and non-bonded ($U_{\textrm{non-bonded}}$) interactions as:

\begin{equation}\label{equation:totenergy}
    U_{\textrm{total}} = U_{\textrm{bonded}} + U_{\textrm{non-bonded}} = U_{\textrm{bond}} + U_{\textrm{angle}} + U_{\textrm{torsion}} + U_{\textrm{vdW}} + U_{\textrm{electrostatics}}, 
\end{equation}

where the bonded interactions are defined by the harmonic potential, the angle potential and the torsion potential, shown in 
\begin{equation}\label{equation:harmonic}
  U_{\textrm{bond}}=\sum_{\textrm{bonds}}{} k_{\textrm{bond}}(r-r_0)^2,
\end{equation}
where $r$ is the bond length, $r_0$ is the equilibrium bond length, and the spring constant $k_{\textrm{bond}}$. The sum runs over all paired atoms. 

\begin{equation}\label{equation:angle}
  U_{\textrm{angle}}=\sum_{\textrm{angles}}{} k_{\theta}(\theta-\theta_0)^2,
\end{equation}
where $\theta$ is the bond length, $\theta_0$ is the equilibrium bond length, and the spring constant $k_\theta$. The sum runs over all angles formed by three consecutive linked atoms. 

\begin{equation}\label{equation:torsions} 
U_{\text{torsion}} \;=\; \sum_{\text{torsions}} \sum_{n} \frac{1}{2} V_n \left[ 1 + \cos\!\big(n\phi - \gamma\big) \right],
\end{equation}
where $\phi$ is the dihedral angle, $V_n$ is the torsion barrier height, $n$ is the periodicity and $\gamma$ is the phase angle. The sum runs over all dihedral planes. 

Non-bonded interactions consist of the sum of the hydrophobic interaction and electrostatic interaction. The hydrophobic interaction is given by the Lenard--Jones 12-6 form potential. This potential is defined as
\begin{equation}\label{eq: VdW potential}
    U_\text{vdW}(r_\textit{i,j}) = \sum_{i<j} 4\varepsilon_{ij} \left[ \left( \frac{\sigma_{ij}}{r_{ij}} \right)^{12} \;-\; \left( \frac{\sigma_{ij}}{r_{ij}} \right)^{6} \right],
\end{equation}

where $r_{ij}$ is the distance between atoms $i$ and $j$, $\varepsilon_{ij}$ is the depth of the potential well and $\sigma_{ij}$ is the distance at which potential is zero.

Here $\sigma_{ij}$ is the pair-of-beads diameter defined from the individual diameter ($\sigma_i$ and $\sigma_j$) assuming the Lorentz-Berthelot mixing rules (i.e., $\sigma_{ij}=(\sigma_i+\sigma_j)/2$). The interaction parameter $\varepsilon_{ij}$ is defined for each specific interaction thanks to Lorentz--Berthelot as $\varepsilon_{ij}=\sqrt{\varepsilon_i\varepsilon_j}$. 

The electrostatic interactions are described by the Coulomb's law. This potential is defined as
\begin{equation}\label{eq:coulomb_potential}
    U_\text{electrostatic} = \sum_{i<j} \frac{q_{i}q_{j}}{4\pi\varepsilon_0r_{ij}},
\end{equation}
where $q_{i},\:q_{j}$ are the partial atomic charges and $\varepsilon_0$ is the vacuum permittivity. Cutoff for this interaction is 9 \AA. Long-range electrostatic interactions were treated using the particle mesh Ewald (PME) method.

All simulations with this force field were carried out using the molecular dynamics Gromacs\_mpi 2023.\cite{gromacsAA}

Topology files, mdp files and some initial configurations, where all parameters required to describe the intermolecular interactions, are available in this \href{https://github.com/Alex-castro-quim/Glyphosate_crystallization}{GitHub link for the repository}.

\section{FEP+ and OPLS4\label{FEP-and-OPLS}} 

For FEP+ (Free Energy Perturbations) calculus with OPLS4\cite{lu2021opls4AA}.
We created systems composed of 15000 solvent molecules, 1 solute molecule and the adequate quantity of glycine impurities for the different concentrations. 
20 different seeds of the system were built to increase sampling and gather convergence statistics using the “tangled chain” option from Schr\"{o}dinger Materials Studio.

For the FEP calculations, we use the default set up which creates 12 $\lambda$ windows.
$\lambda$ is a coupling parameter that modifies the interaction energy of the system.
A $\lambda$ window is the simulation performed at a specific intermediate value of $\lambda$, representing a partially transformed state between the initial and final systems.
At $\lambda = 0$, the system corresponds to the initial non-solvated state, while at $\lambda = 1$ it corresponds to the final solvated state.
In this case we modify the Van der Waals and Columbic contributions, meaning that at $\lambda=0$ the interatomic interactions between solute and solvent are switched off, and at $\lambda=1$, the interactions are fully switched on, and intermediate values represent scaled interactions.
In the isobaric-isothermal ensemble ($NpT$) the difference in Gibbs free energy between two systems is given by equation \ref{eq:FEP}\cite{vega2008determinationAA}.

\begin{equation} \label{eq:FEP}
G(N,p,T,\lambda=1) = G(N,p,T,\lambda=0) 
+ \int_{\lambda=0}^{\lambda=1} 
\left\langle \frac{\partial U(\lambda)}{\partial \lambda} \right\rangle_{N,p,T,\lambda} 
\, d\lambda.
\end{equation}

where G is the Gibbs free energy, and U is the total energy.
%Lambda hopping is activated so that exchange between these $\lambda$ simulations is attempted every 1.2 ps.

To relax the system, we run a ladder of different equilibration simulations, as recommended by Schr\"{o}dinger, in the solvation protocol\cite{schrodinger2023-1AA}:

\begin{enumerate}
    \item NVT Brownian Dynamics for 0.1 ns at 10.0 K and a timestep of 1 fs
    \item Molecular Dynamics 0.024 ns/NVT/300.0 K/1 fs
    \item Molecular Dynamics 0.24 ns/NVT/700.0 K/1 fs
    \item Molecular Dynamics 0.024 ns/NPT/1.01325 bar/300.0 K/1 fs
    \item Molecular Dynamics 0.24 ns/NPT/1.01325 bar/300.0 K/2 fs
    \item Molecular Dynamics 20 ns/NPT/1.01325 bar/300.0 K/2 fs
    \item Molecular Dynamics 10 ns/NPT/1.01325 bar/300.0 K/2 fs
\end{enumerate}

Temperature is kept constant using a Nosé-Hoover chain thermostat\cite{nose1984unifiedAA} with a relaxation time of 1 ps and the pressure is kept constant with an MTK barostat\cite{martyna1994mtkAA} with a relaxation time of 2 ps. Equations of motion are integrated using RESPA\cite{tuckerman1992reversibleAA}, and short-range columbic interactions are cut at 0.9 nm.

Free energies are estimated with the BAR (Bennett Acceptance Ratio) method, using the fep\_analysis script from Schr\"{o}dinger tool set\cite{schrodinger2023-1AA}, which performs Hamiltonian integration calculations and yields the solvation free energy.
The BAR method estimates free energy differences by combining forward and reverse energy samples between adjacent $\lambda$ windows, providing a low-variance calculation of the total free energy change.

\section{\label{sec:methods}Calculation of solubilities via Direct Coexistence simulations}

The determination of solubilities was performed using the Direct Coexistence (DC) method. The crystal is placed in a square prismatic elongated box where an aqueous solution of 8000 water molecules is put in contact with the liquid phase, where the molecule whose solubility we aim to calculate is disolved. The long side of the box is perpendicular to the interfaces. 

Molecular Dynamic simulations were performed using Gromacs 2023 simulation package\cite{gromacsAA}.
All runs were performed at 1 bar constant pressure and 300K constant temperature. 
DC simulations were performed in the $NpT$ ensemble using an anisotropic Parrinello-Rahman barostat\cite{ParrinelloRahman1981AA} with a relaxation time of 10 ps and a V-rescale thermostat\cite{Bussi2007AA} with a relaxation time of 0.1 ps. 
The time step for the Leap-frog algorithm\cite{hockney1974quietAA} was 2 fs.

Through the study of density profiles obtained with the analysis tool \emph{gmx density} we can identify different regions: the glyphosate crystal, the interface between the aqueous solution and the crystal and the aqueous solution bulk. 
The bulk of the aqueous solution is the region from which we obtain the partial densities to calculate the solubility through equation \ref{equation:solubility_supp}.

\begin{equation}\label{equation:solubility_supp}
    m_{glyphosate}=\frac{\rho_y^{glyphosate}}{\rho_y^{H_2O}\cdot M_{glyphosate}},
\end{equation}

\begin{figure*}[hbt!]
	\centering
	\includegraphics[width=0.99\linewidth]{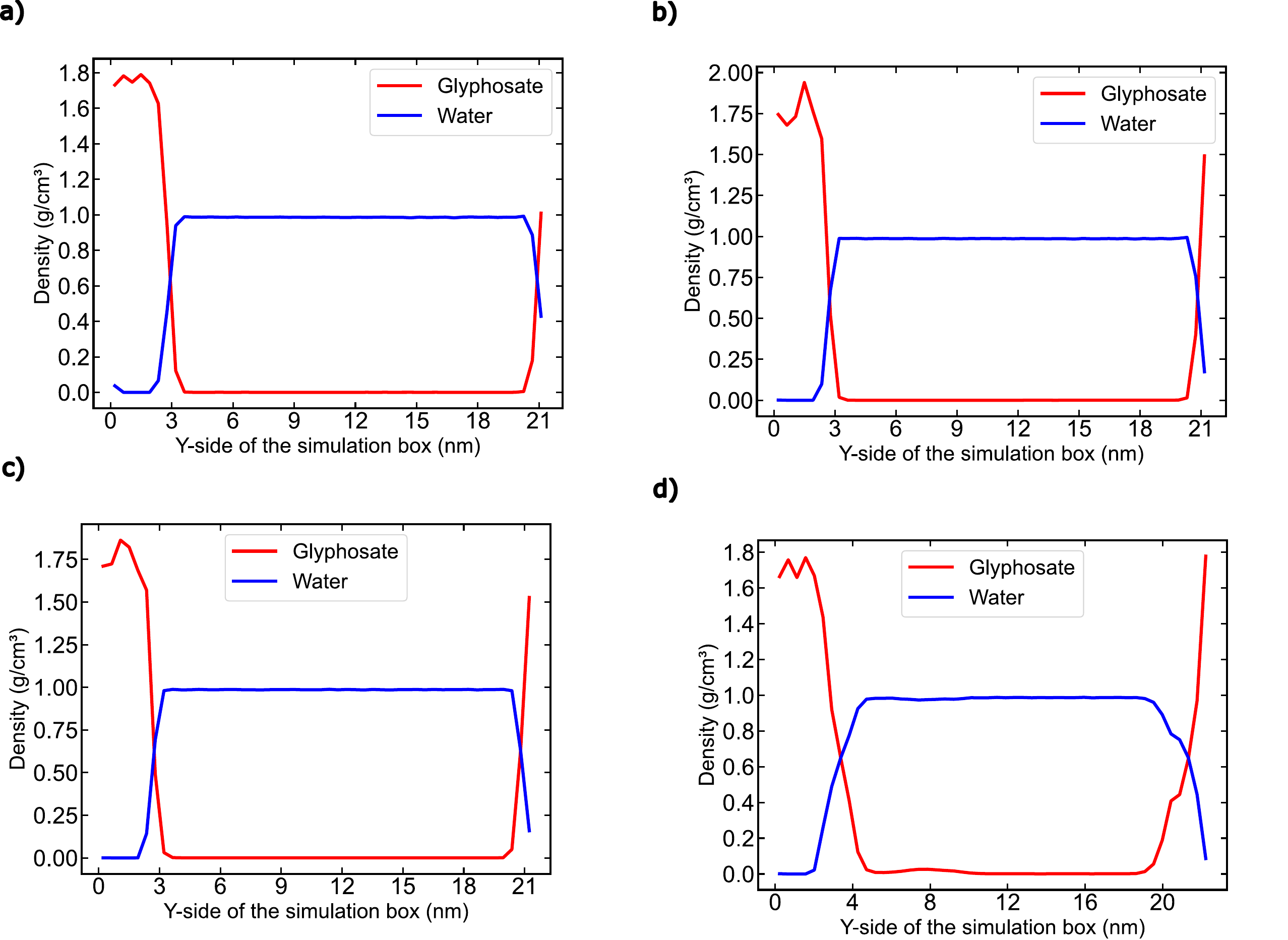}
	\caption{\textbf{All the non vacancy density profiles.} a)~ $C_0=0\:mol/kg$ b)~ $C_0=0.034\:mol/kg$ c)~ $C_0=0.069\:mol/kg$ d)~ $C_0=0.69\:mol/kg$}
 \label{noV}
\end{figure*}

\begin{figure*}[hbt!]
	\centering
	\includegraphics[width=0.99\linewidth]{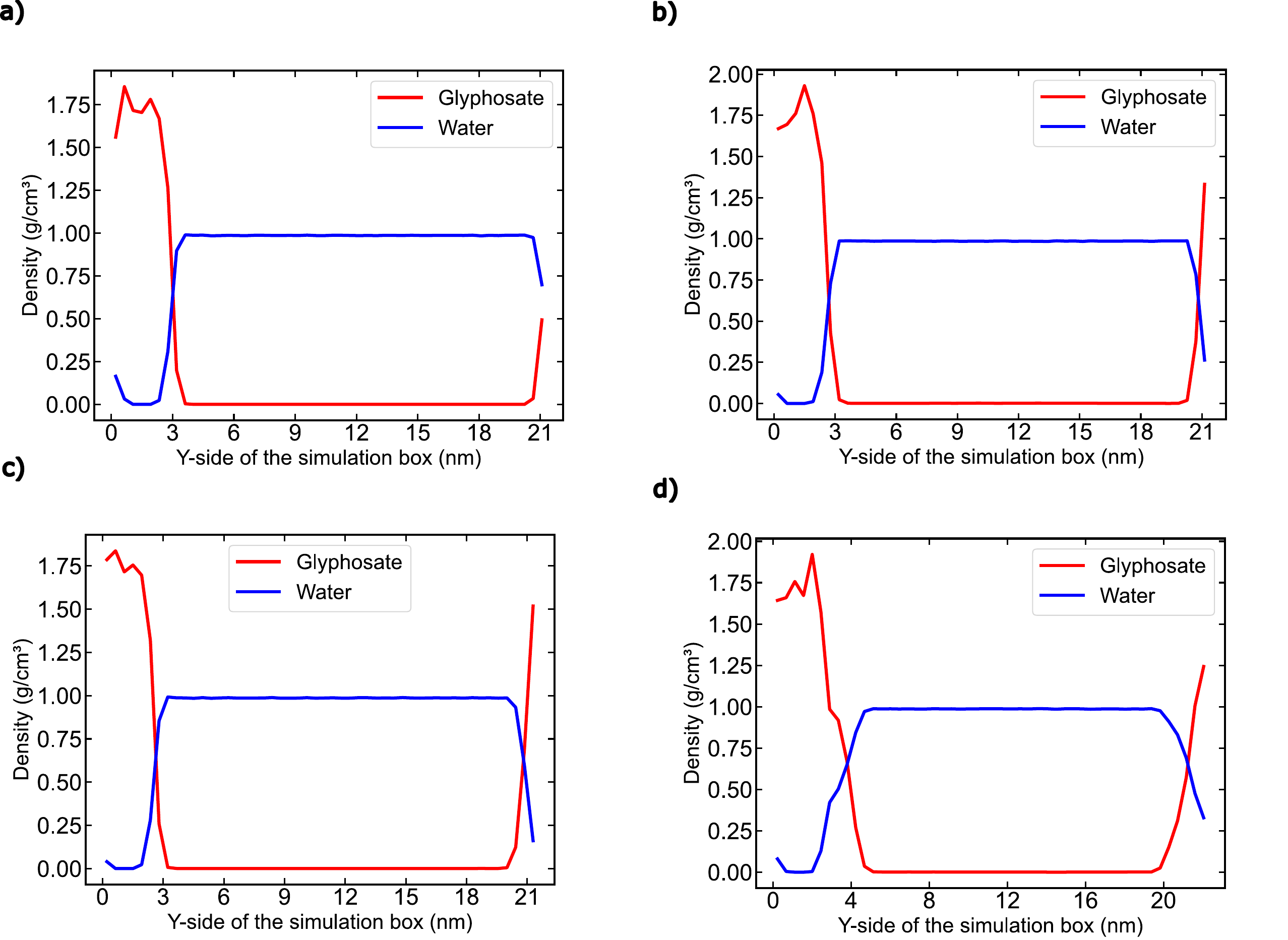}
	\caption{\textbf{All the vacancy density profiles.} a)~ $C_0=0\:mol/kg$ b)~ $C_0=0.034\:mol/kg$ c)~ $C_0=0.069\:mol/kg$ d)~ $C_0=0.69\:mol/kg$}
 \label{V}
\end{figure*}

The density profiles for all studied systems are shown in figures \ref{noV} and \ref{V}

The results from applying equation \ref{equation:solubility_supp} to the density profiles are summarized in table \ref{tab:res}

\begin{table}[h]
\centering
\begin{tabular}{|c|c|}
\hline
Simulation details                 & $m_{gly}\:(mM)$  \cr
\hline
0 glyphosate molecules added in a non-vacancy crystal                              & $2.4 \pm 0.5$           \cr
5 glyphosate molecules added in a non-vacancy crystal                     & $3 \pm 2$                     \cr
10 glyphosate molecules added in a non-vacancy crystal                   & $3.1 \pm 0.5$                           \cr
100 glyphosate molecules added in a non-vacancy crystal                   & $7.0 \pm 0.5$                     \cr
\hline
0 glyphosate molecules added in a 10 vacancy crystal            & $1.0 \pm 0.5$                  \cr
5 glyphosate molecules added in a 10 vacancy crystal   & $7.6 \pm 0.5$                    \cr
10 glyphosate molecules added in a 10 vacancy crystal  & $2.1 \pm 0.8$                    \cr
100 glyphosate molecules added in a 10 vacancy crystal & $4.8 \pm 0.5$                     \cr
\hline
\end{tabular}
\caption{Solubility results from the different DC simulations expressed as molarity}
\label{tab:res}
\end{table}

\section{Other interesting systems \label{other-systems}}

\begin{figure*}[hbt!]
	\centering
	\includegraphics[width=0.99\linewidth]{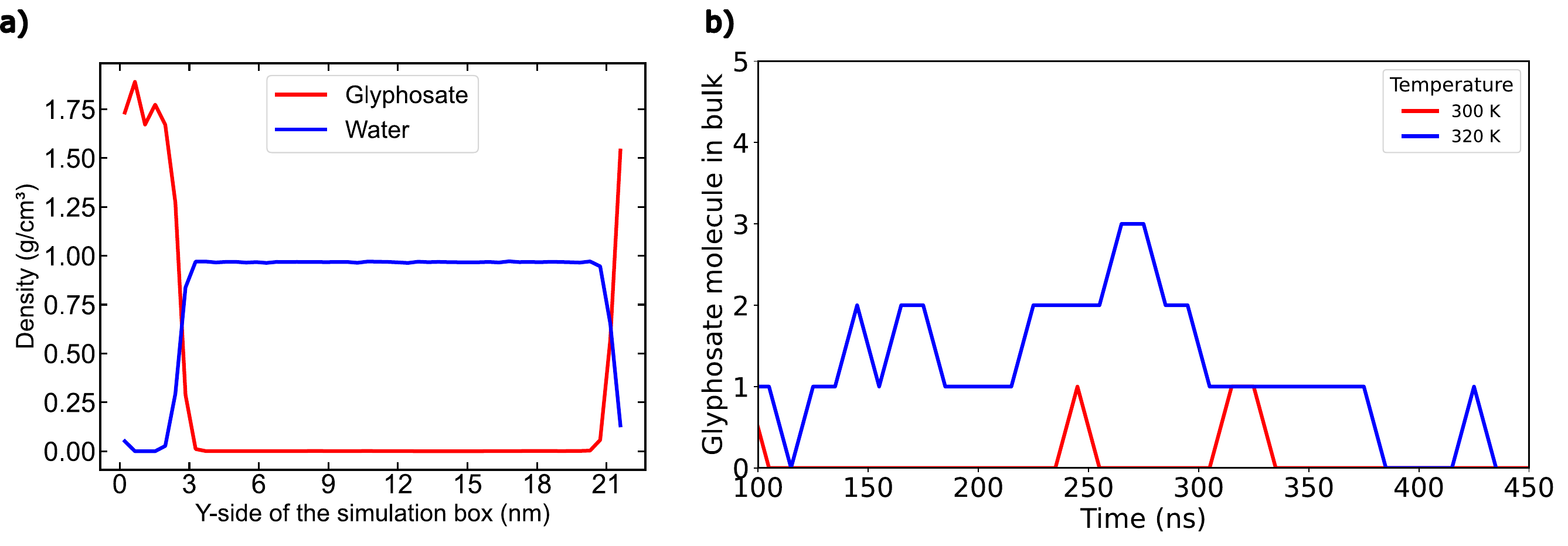}
	\caption{\textbf{Higher temperature experiments} a)~Density profile of a 320 K simulation. b)~Comparison of the number of glyphosate molecules in bulk as a function of time for simulations conducted at 320 K and 300 K.}
 \label{T_image}
\end{figure*}

Additionally we studied how different factors could affect glyphosate solubility.
First, we study the effect of a 20 K increment in our simulations. 
The goal of doing this simulations is to check wether the system is kinetically arrested.
In Fig~\ref{T_image}.a we show the density profile for 320 K, from which we obtained the solubility results using \ref{equation:solubility_supp}.
The comparison at 300 K ($4\pm3\:mmol/kg$) and 320K ($7\pm2\:mmol/kg$) yields interesting results. 
Despite showing the same solubility within the statistical error, we can see in Fig~\ref{T_image}.b that a higher temperature promotes the existence of glyphosate in the solution bulk.
We can also conclude that a 20 K increase is insufficient to produce statistically significant differences in our system, although it is enough to suggest a possible trend.

Moreover, we performed DC simulations for a system with a smaller cross section, implying a smaller interface area. 
Here we might find finite size effects to be relevant.
Our simulations confirm this fact, as we find that there is no glyphosate molecule liberated from the crystal to the bulk of the aqueous solution, opposed to what we observe in Fig. \ref{noV}.

\section{Glycine insertion \label{glycine}}

\begin{figure*}[hbt!]
	\centering
\includegraphics[width=0.75\linewidth]{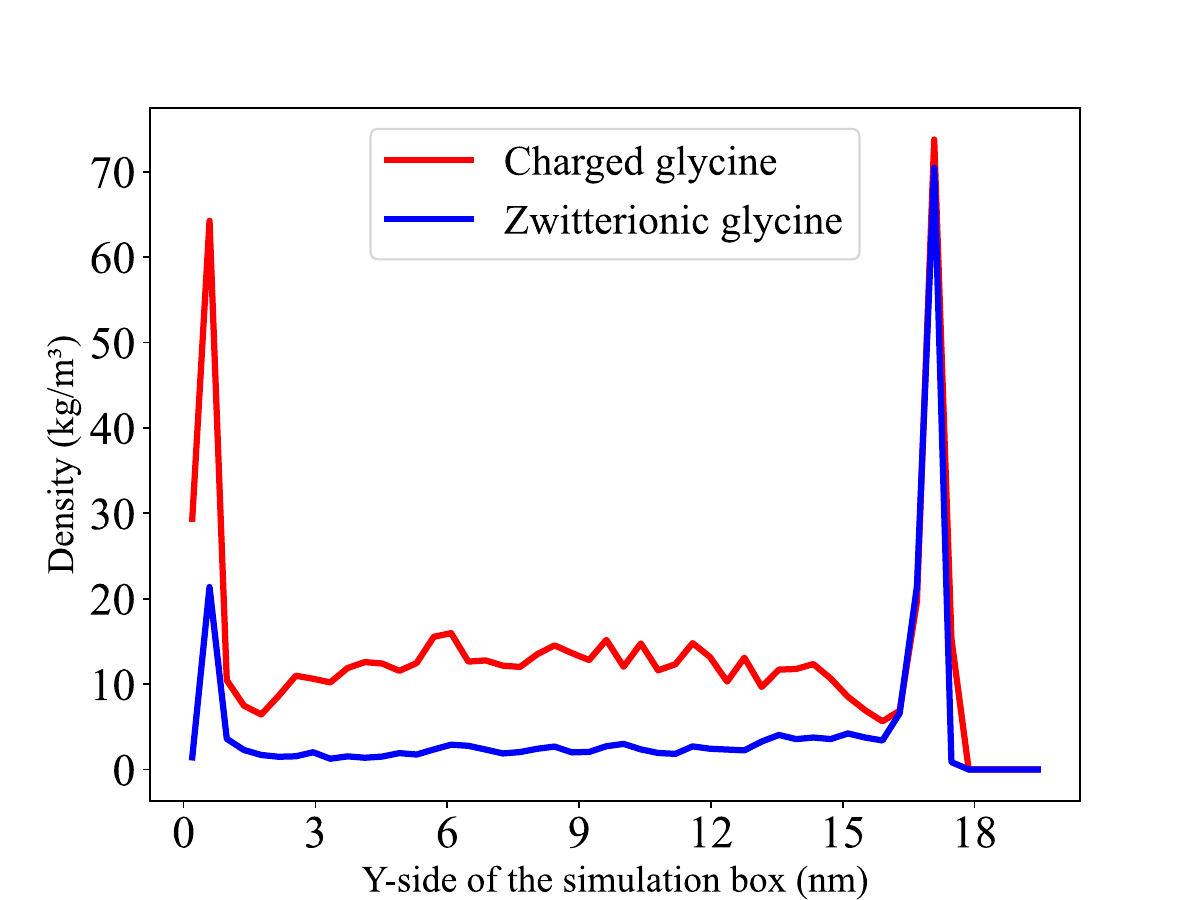}
	\caption{Density profile of the different glycine species.}
 \label{supp:Glycine}
\end{figure*}

Glycine, the ubiquitous impurity present in glyphosate synthesis, must be taken into account in our simulations.
As such, we need to study glycine at the experimental conditions used in glyphosate synthesis.
At pH = 1.9, glycine shows two predominant forms, charged and zwitterionic \cite{yu2002glycineAA}, therefore, we took both species into account as separate entities for our forcefield, maintaining the experimentally observed ratio of 25:75 at pH = 1.9 for zwitterionic:charged glycine.
We analyzed the different glycine species individually to examine whether their form (zwitterionic or charged) is related to their behavior.
The results are shown in Figure \ref{supp:Glycine}, where we can observe how both species have maxima, which are located at the interface between glyphosate and the aqueous solution, exhibiting a similar behaviour.

\begin{figure*}[hbt!]
	\centering
\includegraphics[width=1\linewidth]{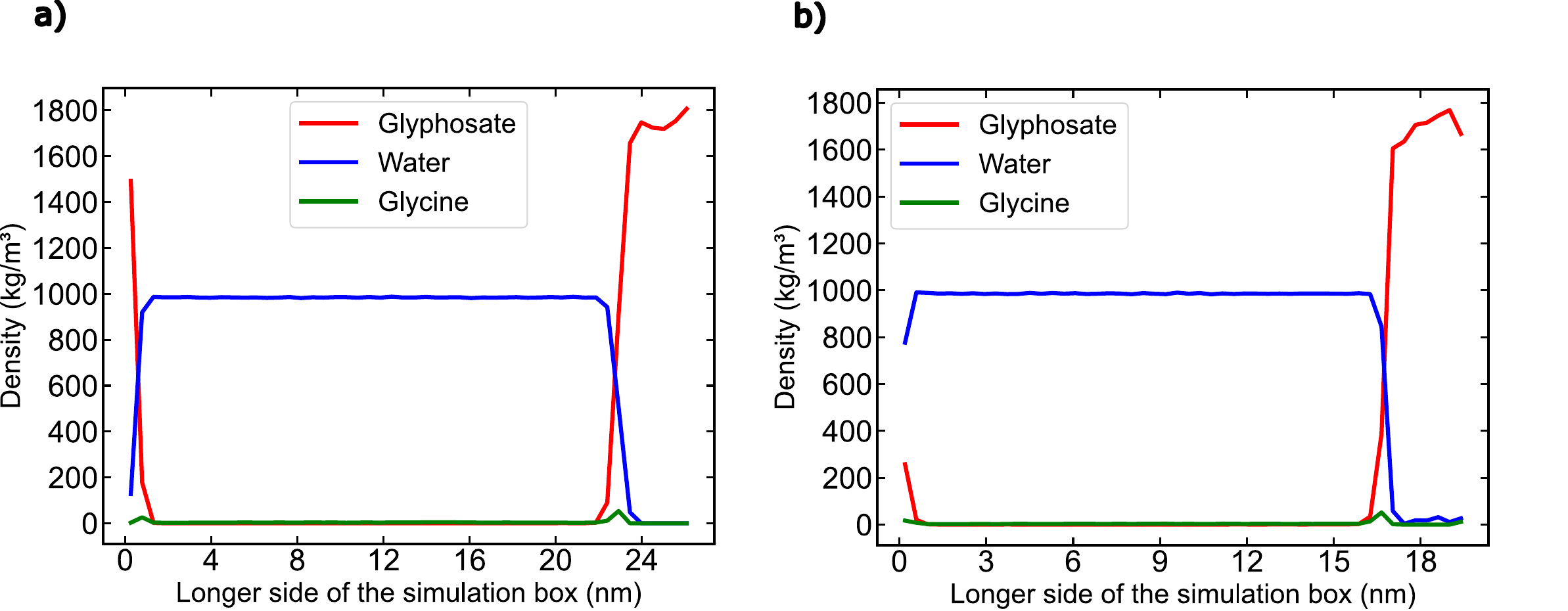}
	\caption{\textbf{Glycine density profiles.} a)~Z-axis simulation with equimolar conditions b)~Y-axis simulation with equimolar conditions}
 \label{supp:Glycine2}
\end{figure*}

The main difference between them that we observe in Figure \ref{supp:Glycine} is that although zwitterionic glycine is present in smaller proportion, it exhibits a stronger tendency to adhere to the crystal face compared to the charged glycine.
To study this observation we calculated the proportion of glycine at the interface and within the bulk.
The results of this analysis are presented in Table \ref{tab:proportion}, showing quantitative differences in the percentages and indicating that zwitterionic glycine forms a stronger interfacial coating.
One explanation for this phenomena is the possibility of charged glycine of inducing dipoles in water, therefore observing more affinity for water than its zwitterionic counterpart.
The density profiles for the different glycine experiments are shown in \ref{supp:Glycine2}

\begin{table}[h]
\centering
\caption{Proportion of glycine by location}
\label{tab:proportion}
\begin{tabular}{|c|c|c|}
\hline
                     & \% Glycine at the interface & \% Glycine in bulk \\
\hline
Charged glycine      & 33.1                    & 66.9               \\
Zwitterionic glycine & 55.5                    & 45.5              \\
\hline
\end{tabular}
\end{table}

The density profiles for the different glycine experiments are shown in \ref{supp:Glycine2}

\section{FEP+} 

\begin{figure*}[hbt!]
	\centering
	\includegraphics[width=0.99\linewidth]{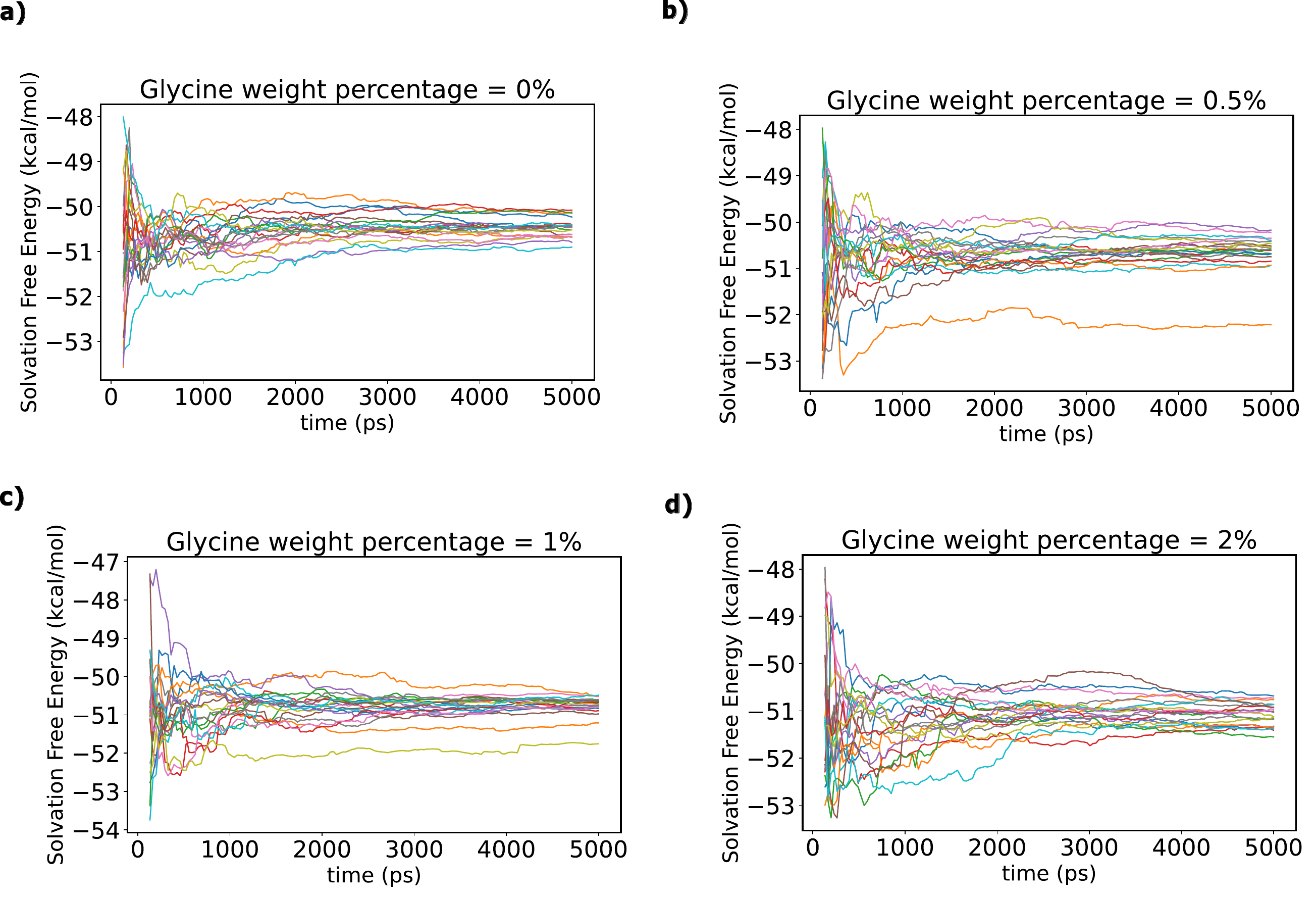}
	\caption{Solvation free energy \emph{vs} time for each of the 20-seeds (depicted with different colours) for each glycine concentration used.}
 \label{FEP_image}
\end{figure*}

The systems studied were comprised of the molecules shown in Table \ref{tab:Fep_sys}. In Figure \ref{FEP_image} we show the time evolution of the solvation free energy for the different seeds, which are depicted with different colours.

\begin{table}[h]
\centering
\caption{Number of molecules in the different systems studied}
\label{tab:Fep_sys}
\begin{tabular}{|c|c|c|c|c|c|}
\hline
\%wt glycine & $H_2O$ & glyphosate molecules & charged glycine & zwitterion glycine & total glycine \\
\hline
0            & 15000  & 1                    & 0               & 0                  & 0             \\
0.5          & 15000  & 1                    & 14              & 4                  & 18            \\
1            & 15000  & 1                    & 27              & 9                  & 36            \\
2            & 15000  & 1                    & 55              & 18                 & 73           \\
\hline
\end{tabular}
\end{table}

\section*{Error calculation}

In this section we will discuss how errors were obtained.
For the individual simulations shown in \ref{tab:res}, we used the propagation of errors across \ref{equation:solubility_supp}, obtaining the expression shown in \ref{equation:sol_error}

\begin{equation}\label{equation:sol_error}
\Delta m_{\text{glyphosate}}
= m_{\text{glyphosate}} \cdot \left[
\frac{\Delta \rho_{y}^{\text{glyphosate}}}{\rho_{y}^{\text{glyphosate}}}
+ \frac{\Delta \rho_{y}^{\text{H}_2\text{O}}}{\rho_{y}^{\text{H}_2\text{O}}}
\right],
\end{equation}

For final solubility results we used the standard statistical error obtained with \ref{eq:stat_err}.

\begin{equation}\label{eq:stat_err}
\sigma = \sqrt{\frac{1}{N} \sum_{i=1}^{N} (x_i - \mu)^2},
\end{equation}

where $x_i$ is the $i$-th observation in the dataset, $N$ is total number of observations in the population, $\mu$ is the population mean and $\sigma$ is population standard deviation

For the FEP+ error, it is first necessary to define how the BAR method calculates the variance. The variance is determined from the curvature of the log-likelihood at the optimum, as shown in Eq.~\ref{eq:barerror}.

\begin{equation}
\label{eq:barerror}
\sigma^2(\Delta G) \approx 
\left[- \frac{\partial^2}{\partial (\Delta G)^2} \ln L(\Delta G) \Big|_{\Delta G = \Delta G^*} \right]^{-1},
\end{equation}

where $\Delta G^*$ is the optimal solution to the BAR equation and $L$ is the likelihood function for the free energy difference $\Delta G$, defined in \ref{eq:likelihood}

\begin{equation}
\label{eq:likelihood}
L(\Delta G) = \prod_{i=1}^{N_A} f\Big(-\beta(\Delta U_i^{A \to B} - \Delta G)\Big)
\prod_{j=1}^{N_B} f\Big(\beta(\Delta U_j^{B \to A} - \Delta G)\Big),
\end{equation}
where $f(x) = \frac{1}{1 + e^x}$ is the Fermi function, $\beta = 1/k_B T$, and $N_A, N_B$ are the number of samples from states $A$ and $B$ respectively.

Taking the logarithm in \ref{eq:likelihood}, the log-likelihood is:
\begin{equation}
\label{eq:loglikelihood}
\ln L(\Delta G) = \sum_{i=1}^{N_A} \ln f\Big(-\beta(\Delta U_i^{A \to B} - \Delta G)\Big)
+ \sum_{j=1}^{N_B} \ln f\Big(\beta(\Delta U_j^{B \to A} - \Delta G)\Big).
\end{equation}

The BAR free energy estimate $\Delta G^*$ is obtained by maximizing the log-likelihood.
Equation~\ref{eq:barerror} provides the standard error of the free energy difference computed by BAR. 

\section{Radial dsitribution functions}

\begin{figure*}
    \centering
    \includegraphics[width=0.95\linewidth]{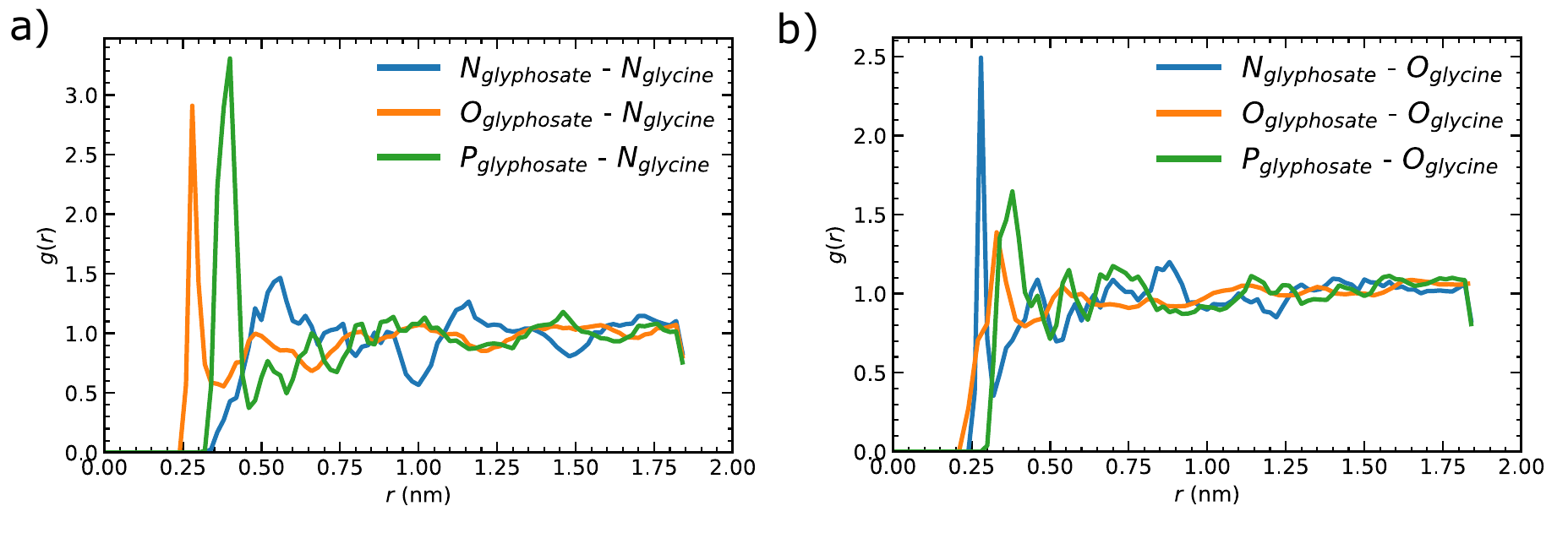}
    \caption{\textcolor{black}{a) Radial distribution functions g(r) between the N, O, and P atoms of glyphosate and the N atom of glycine as a function of distance r. b) Radial distribution functions, g(r), between the N, O, and P atoms of glyphosate and the O atom of glycine as a function of distance r. }}
    \label{figrdfs}
\end{figure*}

\textcolor{black}{In order to characterize which are the most relevant glycine-glyphosate interactions driving the increase in solubility, we compute the radial distribution function (g(r)) between the different heavy atoms of both molecules. The results are summarized in Figure \ref{figrdfs}. From the radial distribution functions between glycine's nitrogen with gluphosate heavy atoms, we conclude that the most remarkable interaction is that between glyphosate's oxygen (negative dipole moment of the molecule) with glycine's nitrogen (positive dipole moment of the molecule), which is expected given the high amount of oxygen atoms (5 in every molecule) per glyphosate, all of which have a negative dipole moment. The following peaks (P$_{Glyphosate}$-O$_{Glycine}$ and N$_{Glyphosate}$-O$_{Glycine}$) appear naturally due to the proximity of these atoms in the neighbouring molecules, binded by oppositely charged groups. From the radial distribution functions between glycine's oxygen and glyphosate heavy atoms, we do not clearly observe a first coordination shell between any pairs of atoms, highlighting the stabilizing role of the O$_{Glyphosate}$-N$_{Glycine}$ pair.
}

\clearpage

\section{Molecular Dynamics system size}

In tables S4 and S5 we explicitely state the number of molecules of each species for each simulation mentioned in the main text.

\begin{table}[h!]
\centering
\caption{Number of molecules for glyphosate simulations.}
\label{tab:glyphosate_only}
\begin{tabular}{|l|cccccc|}
\hline
 & \multicolumn{2}{c}{\textbf{Crystal simulations}} 
 & \multicolumn{4}{c|}{\textbf{Bulk concentrations}} \\
\cline{2-3} \cline{4-7}
\textbf{Component} 
& No vacancies 
& Vacancies 
& 0\,m 
& 0.034\,m 
& 0.069\,m 
& 0.69\,m \\
\hline
Water molecules      & --    & --    & 8000 & 8000 & 8000 & 8000 \\
Glyphosate molecules & 256   & 246   & 0    & 5    & 10   & 100  \\
\hline
\end{tabular}
\end{table}

\begin{table}[h!]
\centering
\caption{Number of molecules for glycine simulations.}
\label{tab:glycine}
\begin{tabular}{|l|ccc|}
\hline
 & \textbf{Crystal simulation} & \multicolumn{2}{c|}{\textbf{Bulk concentrations}} \\
\cline{3-4}
\textbf{Component} 
& No vacancies 
& 0.5\,\% 
& 2\,\% \\
\hline
Water molecules      & --    & 8000 & 8000 \\
Glyphosate molecules & 256   & 10   & 10   \\
Glycine molecules    & --    & 10   & 40   \\
\hline
\end{tabular}
\end{table}

\bibliographystyle{ieeetr}

\end{document}